\documentclass[referee]{gji}
\usepackage{graphics}

\title[Absolute palaeointensity from the Kerguelen
Archipelago]{Absolute palaeointensity of Oligocene (28-30 Ma) lava
  flows from the Kerguelen Archipelago (southern Indian Ocean)}

\author[G. Plenier, P. Camps, R.S. Coe and M. Perrin]{G. Plenier$^1$,
  P. Camps$^2$\thanks{Corresponding author. Tel: +33 467 14 39 38;
    Fax: +33 467 14 36 03. e-mail address: camps@dstu.univ-montp2.fr},
  R.S. Coe$^3$ and M. Perrin$^2$\\
  $^1$ Laboratoire G\'{e}ophysique, Tectonique et S\'edimentologie,
  CNRS and ISTEEM, Universit\'e
  Montpellier 2\\ case 060, 34095 Montpellier Cedex 05, France\\
  $^2$ Laboratoire de Tectonophysique, CNRS and ISTEEM, Universit\'e Montpellier 2 case 049,\\
  34095 Montpellier Cedex 05, France\\
  $^3$ Earth Science Dept, University California, Santa Cruz, CA
  95064,USA}

\date{Submitted to {\it Geophysical Journal International}\\
  March XX, 2003 } \pagerange{\pageref{firstpage}--\pageref{lastpage}}
\pubyear{2003}

\begin{document}
\label{firstpage}

\maketitle

\begin{summary}
  We report palaeointensity estimates obtained from three Oligocene
  volcanic sections from the Kerguelen Archipelago (Mont des Ruches,
  Mont des Temp\^{e}tes, and Mont Rabouill\`{e}re). Of 402 available
  samples, 102 were suitable for a palaeofield strength determination
  after a preliminary selection, among which 49 provide a reliable
  estimate.  Application of strict {\textit{a posteriori}} criteria
  make us confident about the quality of the 12 new mean-flow
  determinations, which are the first reliable data available for the
  Kerguelen Archipelago.  The Virtual Dipole Moments (VDM) calculated
  for these flows vary from 2.78 to 9.47 10$^{22}$ Am$^2$ with an
  arithmetic mean value of 6.15$\pm$2.1 10$^{22}$ Am$^2$.  Compilation
  of these results with a selection of the 2002 updated IAGA
  palaeointensity database lead to a higher (5.4$\pm$2.3
  10$^{22}Am^2$) Oligocene mean VDM than previously
  reported~\cite{Avto01,Riisager99}, identical to the 5.5$\pm$2.4
  10$^{22}Am^2$ mean VDM obtained for the 0.3-5 Ma time window.
  However, these Kerguelen palaeointensity estimates represent half of
  the reliable Oligocene determinations and thus a bias toward higher
  values. Nonetheless, the new estimates reported here strengthen the
  conclusion that the recent geomagnetic field strength is anomalously
  high compared to that older than 0.3 Ma.
\end{summary}


\section{Introduction}
Numerous studies have been carried out to increase our knowledge of
geodynamo physics, but even so we still do not know the detailed
mechanism of the generation of the Earth's magnetic field, and even
less about the processes that produce secular variation, excursions
and reversals (e.g., Jacobs~\shortcite{Jacobs94}; Merrill \&
McFadden~\shortcite{Merrill99}). To better understand the geomagnetic
field we need to be able to go back in time in order to observe its
changes and to obtain long-term global characteristics.  This is
possible with some rocks which recorded the Earth's palaeomagnetic
field during their formation. Volcanic rocks, in particular, furnish a
global knowledge of the geomagnetic field because they contain
information on both the direction (inclination and declination) and
the strength of the palaeofield.  However, the most reliable methods
for absolute palaeointensity determination, Thellier \&
Thellier~\shortcite{Thellier59} and its modified version proposed by
Coe~\shortcite{Coe67a}, are time consuming because of the strict
conditions which have to be checked to validate the determinations.
Moreover, many volcanic rocks turn out to be unsuitable for
palaeointensity determination.  For these reasons, reliable
palaeointensity data are difficult to obtain and are particularly
rare. Only 1.5 determinations per million years between 0 and 300
Ma~\cite{Selkin00} are available when combining the updated IAGA 1999
data-set and the Scripps submarine basaltic glass databases.  It is
obvious that more palaeointensity data are needed for a
better understanding of the geomagnetic dynamo.\\
This study on three basaltic sections sampled in the Kerguelen
Archipelago (49.9\degr S, 70\degr E) aims to estimate more accurately
the palaeomagnetic strength of the geomagnetic field in the 28-30 Ma
time interval recorded by these lavas. It thus complements the
palaeomagnetic directions recently published on the same sections
~\cite{Plenier02}. By combining these new determinations with selected
records issued from preexisting palaeomagnetic databases (Tanaka et
al.~\shortcite{Tanaka95}, updated by Perrin \&
Shcherbakov~\shortcite{Perrin97} and Perrin et
al.~\shortcite{Perrin98a}) we will propose a more robust estimation of
the Oligocene palaeofield strength and we will discuss the long-term
characteristics of the geodynamo.

\section{Geology and sampling}
\begin{figure}
  \includegraphics{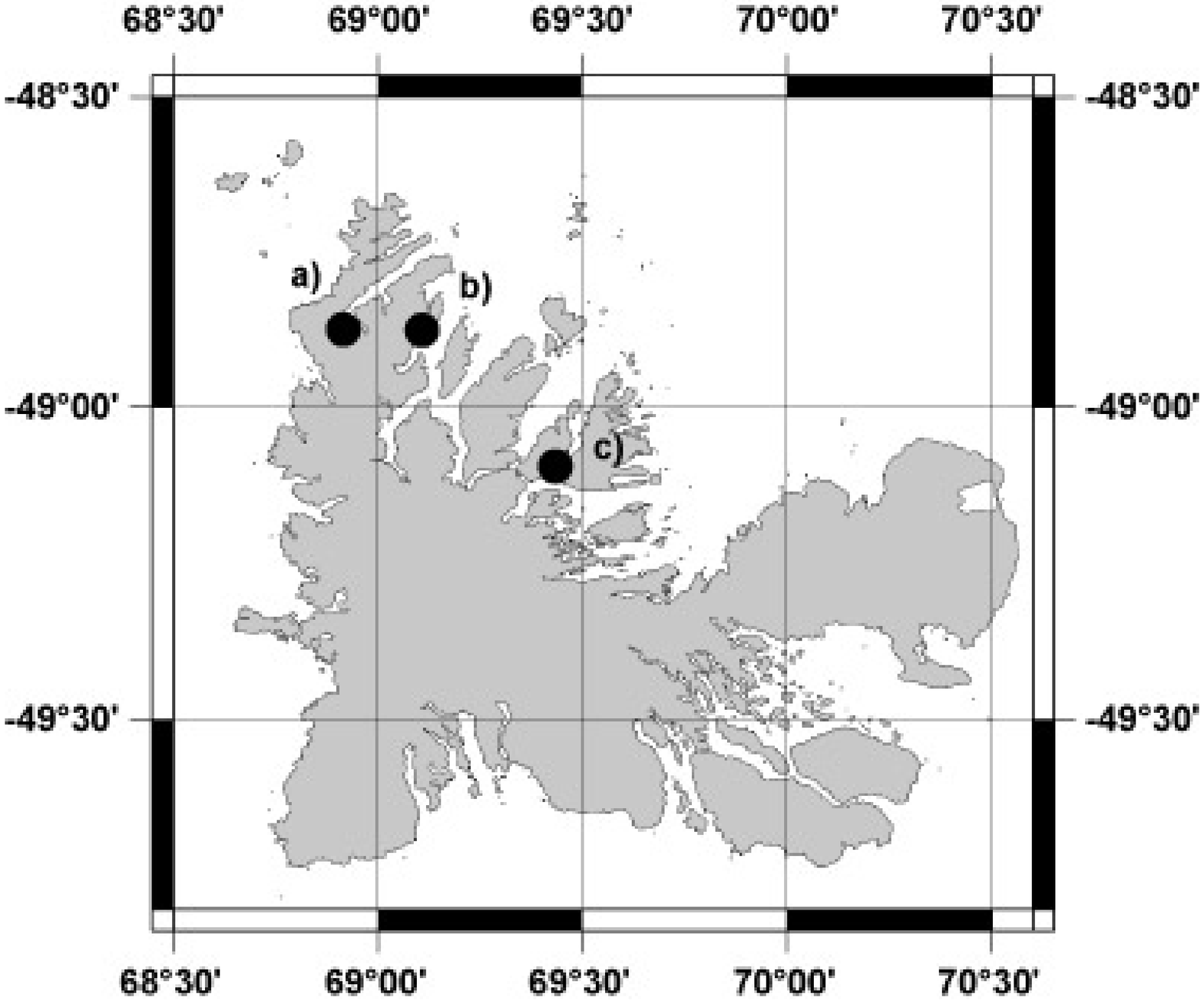}
\caption{Location of the studied sections: a) Mont des Ruches : 18 flows (48\degr52'18''S, 68\degr54'48''E), b) Mont des
  Temp\^etes : 20 flows (48\degr52'50''S, 69\degr06'37''E), c) Mont
  Rabouill\`ere : 19 flows (49\degr05'25''S, 69\degr26'25''E).}
\label{sampl}
\end{figure}

The Kerguelen Archipelago lies on the northern part of the
Kerguelen-Gaussberg Plateau (southern Indian Ocean). This archipelago
is the subaerial continuation of Kerguelen hotspot volcanism for the
last 30 Ma ~\cite{Yang98,Weis98,Nicolaysen00}. The lava flows form the
tabular reliefs (400 to 900 m high) observed today after glacial
erosion, and represent more than 85\% of the archipelago
surface~\cite{Giret90}. The rest of the archipelago is composed of
intrusions (gabbro, granite, and syenite) issued from Kerguelen plume
melts~\cite{Weis94} and quaternary glacial sediments. We studied
paleomagnetic cores collected along three vertical sections of
Kerguelen basalt at Mont des Ruches, Mont des Temp\^{e}tes, and Mont
Rabouill\`{e}re sections (Fig.~\ref{sampl}). For the reasons developed
in Plenier et al.~\shortcite{Plenier02}, the palaeomagnetic sections
do not correspond exactly to the previously-dated
sections~\cite{Yang98,Nicolaysen00,Doucet02}, but they appear to be
correlated. Usually, we drilled seven samples in each successive lava
flow using a gas-powered drill and oriented them with both solar
sightings and magnetic compass with a clinometer. We took care to
sample the bottom part of the least altered flows, and as far away as
possible from intrusions.

\section{Rock magnetism and sample selection}
For field intensities comparable to those of the Earth's magnetic
field (few tens of $\mu$T), there is a proportionality between the
Thermo-Remanent Magnetization (TRM) intensity measured at 20\degr C
and the strength of the ancient magnetic field present during cooling
through the blocking temperatures for almost all natural rocks. Thus,
for some particular rocks cooling in the geomagnetic field during
their formation, it is possible to estimate the palaeomagnetic field
strength recorded by comparing their Natural Remanent Magnetization
(NRM) with an artificial TRM acquired in the laboratory under a known
ambient field.  However, the coefficient of proportionality depends on
grain size, shape distribution, and blocking temperatures as well as
on the amount and type of ferromagnetic material the rock contains.
In addition, this coefficient may have changed since the formation of the
rock or during heating in the laboratory. For this last reason, a
procedure using numerous successive heatings with increasing
temperature steps has been developed~\cite{Thellier59} in order to
limit the field strength estimates to the temperature range preceding
the irreversible magnetic and/or chemical changes in the ferromagnetic
minerals.  The strict conditions to be respected by the samples for
correct palaeointensity determination are the following:

\begin{enumerate}
  
\item {The Characteristic Remanent Magnetization (ChRM) recorded by
    the studied specimen has to be a TRM, acquired at a known epoch in
    the geomagnetic field.}
  
\item {The ChRM should not be disturbed by significant secondary
    magnetizations.}
  
\item {The physical, chemical and crystallographical properties of the
    magnetic minerals must not have changed since the initial TRM was
    acquired nor changed during the successive heatings imposed by the
    experimental method.}
  
\item {The independence and additivity laws of partial-TRM (pTRM)
    enunciated by Thellier~\shortcite{Thellier38} have to be
    satisfied. That is, the total TRM must be equivalent to a sum of
    pTRMs, each associated with its own blocking temperature interval
    and not dependent on the remanence carried in every other
    intervals. This generally means that the magnetic carriers have to
    be single domain (SD) or in favorable cases pseudo-single domain
    (PSD) grains.}
\end{enumerate}
It is obvious that numerous samples can not fulfill these conditions,
thus extensive preliminary studies are necessary to avoid unnecessary
work.

\subsection{Viscosity indices and demagnetizations}
We shown recently~\cite{Plenier02} by means of a positive reversal
test that the ChRM measured from {K}erguelen lava are, generally
speaking, not disturbed by unremoved secondary components and then
that these ChRMs are primary TRMs.  In order to assess the importance
of the secondary component carried by each sample, we analyzed the
results from demagnetization experiments performed previously on
sister
specimens~\cite{Plenier02}.\\
First, we rejected samples for which the angle between the NRM and the
ChRM was greater than 15\degr ~and those contaminated by a resistant
secondary component (e.g., unremoved beyond 20 mT alternating fields
treatment or 300\degr C when samples are demagnetized by heating in
zero field). Likewise, we kept only those flows for which the NRMs of
all samples were well grouped.  60\% of the flows fulfilled these
conditions. For these flows, only the samples with a demagnetization
curve as undisturbed as possible and presenting unblocking
temperatures as high as possible, or
median-demagnetization field of at least 20 mT, were retained.\\
Second, we estimated the capacity of the specimens to gain a secondary
Viscous Remanent Magnetization (VRM) by measuring them first after two
weeks in the ambient magnetic field oriented along the z core axis and
again, after two-week storage in a zero field.  This enables
determination of the viscosity index~\cite{Thellier44} for each
sample, which are reported in Table~\ref{dir}. The viscosity index
corresponds to about 25\% of the VRM acquired in situ since the last
reversal 780 ky ago~\cite{Prevot81}.  We choose the arbitrarily
thresholds for the viscosity index of 10\% as an upper limit for
specimen from intermediate polarity flow and 5\% for the
others~\cite{Prevot85}.

\subsection{Susceptibility at room temperature and k-T curves}
\begin{figure}
  \includegraphics{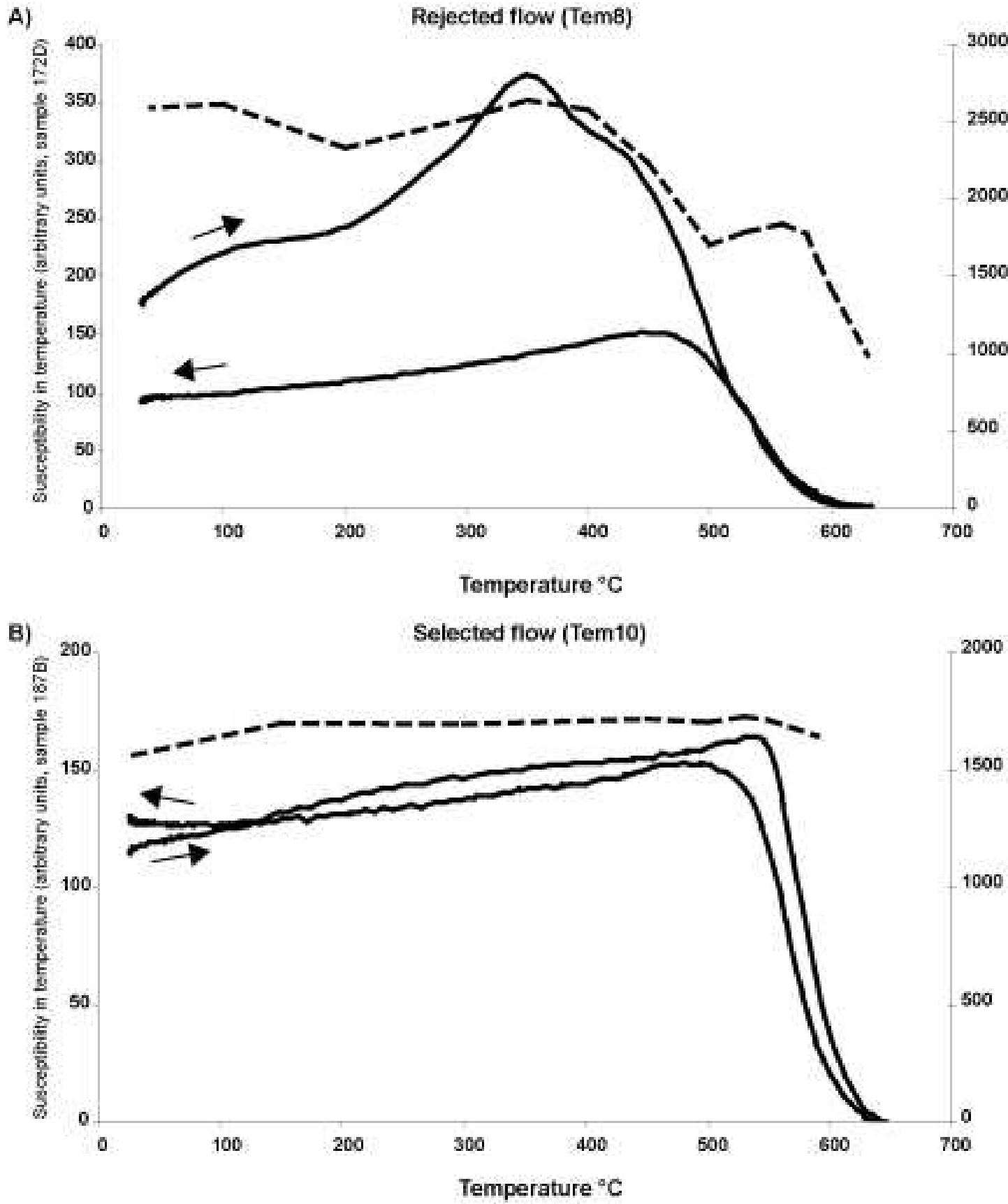}
\caption{Low-field susceptibility versus temperature curves: continuous measurements at temperature k-T (solid line) and room temperature measurements k$_{0}$-T (dashed line) curves of a rejected (A) and a selected (B) flow.} 
\label{kt}
\end{figure}

\begin{figure}
  \includegraphics{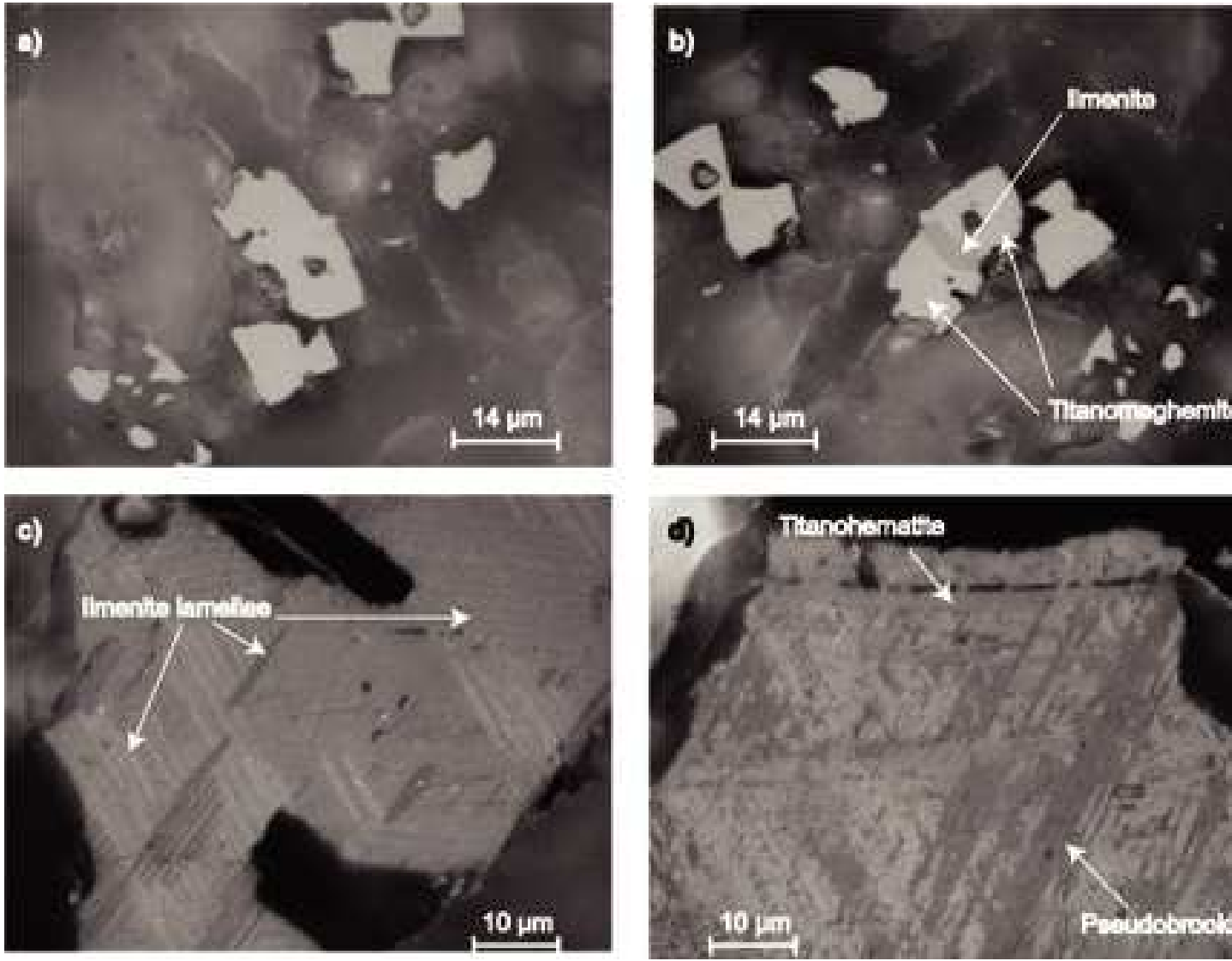}
\caption{Reflected light microphotographs using an oil immersion objective. a),b):
  sample 269b (Tou2) in natural light shown in different orientations in order to illustrated the pleochroic ilmenite associated with a non pleochroic phase which could correspond to a titanomaghemite;
  c),d): two characteristic minerals from sample 234c (Tem16) in
  polarized light. The crossed nicols are at 90\degr} \label{lam}
\end{figure}

To be sure of the thermal stability of the samples during successive
heating in the laboratory, we used the low-field magnetic
susceptibility (k$_{0}$) measured at room temperature after each
thermal demagnetization step performed in air for the sister specimen
previously studied in the palaeodirection
determination~\cite{Plenier02}. A favorable sample for palaeointensity
determination should have a relatively constant k$_{0}$ value during
most of the demagnetization procedure.  However, this is not a
sufficient criterion because thermally unstable samples may
nonetheless display low variations in the susceptibility measured at
room temperature. Thus to complete this approach, we measured
continuously the low-field susceptibility of one sample from each flow
usually during two successive heating-cooling cycles under vacuum, the
first up to 350\degr C and the second up to the Curie temperature (Tc).
Fig.~\ref{kt} presents two representative k-T curves (susceptibility
as a function of temperature) encountered during this study. The first
case (Fig.~\ref{kt}a) illustrates the irreversible and complex
thermomagnetic behaviour observed for almost 60\% of the samples. The
magnetic carriers are interpreted as original titanomagnetite
associated with titanomaghemite, a product of their low temperature
oxidation. We rejected the flows yielding this behaviour because of
their thermal instability.  The second case (Fig.~\ref{kt}b)
illustrates the reversible behaviour observed for the rest of the
samples. The magnetic carriers are low-Ti titanomagnetites probably
produced by high temperature oxyexsolution of the original
titanomagnetites. We considered flows presenting this second
reversible behaviour as suitable for palaeointensity
determination experiments.\\
In order to complement this thermomagnetic investigation, we observed thin
sections from each k-T curve type using an oil immersion objective.
Fig.~\ref{lam}a and ~\ref{lam}b show photomicrographs in natural light
of sample 269b (flow Tou2), which displayed irreversible behaviour. We
saw an isotropic phase sometimes associated with a pleochroic
ilmenite, as illustrated here. Because titanomagnetite and
titanomaghemite are difficult to distinguish under the microscope, it
is hard to conclude about the nature of the isotropic phase. These two
minerals are certainly both present, but the existence of cracks
almost omnipresent in this phase suggest a larger amount of
titanomaghemite. This interpretation agrees well with the irreversible k-T curve. 
Fig.~\ref{lam}c and ~\ref{lam}d show two minerals from
a thermally stable sample (234c, flow Tem16).  They illustrate two
different advanced stages of deuteric oxidation with ilmenite or
titanohematite lamellae exsolved from a residual titanomagnetite
almost entirely altered. Again this observation confirms the
interpretation of the k-T curves.

\subsection{Pilot analysis}
We kept only flows for which at least three samples from different cores
fit all the previously enumerated criteria for the
palaeointensity experiments (32 out of the initial 57). We performed a
pilot analysis with one or two samples from each of these flows in
order to identify the most favorable flows for reliable determinations and
to define the more appropriate demagnetization steps for the next
series. After applying {\textit{a posteriori}} criteria which will be
presented later, 5 flows from the Mont des Ruches section, 9 from the
Mont des Temp\^{e}tes section and 1 from the Mont de la
Rabouill\`{e}re section seemed suitable. Because a sample which did
not pass all the {\textit{a priori}} selection criteria may sometimes
furnish a reliable determination~\cite{Coe67b,Perrin98b}, they should
be regarded as serving only for selecting the most appropriate flows
for palaeointensity determination. For this reason, the remaining
samples from the suitable flows underscored by the pilot analysis have
been incorporated in the following two determination series (68
specimens), including the samples which did not pass all the
{\textit{a priori}} criteria first.

\section{palaeointensity experiments}
\subsection{Experimental procedure}
We used the method of Thellier \& Thellier~\shortcite{Thellier59} in
its classical form to estimate the palaeointensity of the geomagnetic
field. We also followed a sliding pTRM checks
procedure~\cite{Prevot85} every two demagnetization steps in order to
verify that the pTRM capacity remains unaltered when the heating
temperature is progressively increased. Because of the nonlinearity of
acquisition of TRM in multidomain grains~\cite{Levi77}, applying a
laboratory field far from the recorded one may lead to underestimate
the ancient field~\cite{Coe67b,Tanaka84}. Thus, to improve the
quality of data, we took care to apply, along the vertical axis of the
samples, a constant field of 50 $\mu$T (known with a precision of 0.1
$\mu$T), which corresponds approximately to the strength
of the mean Oligocene field for Kerguelen latitude. We also carried
out the heating-cooling cycles under a vacuum better than 10$^{-4}$
mbar to limit possible oxidation during experiments. We performed
demagnetizations up to 580\degr C, with 13 steps ranging from 50 to
10\degr C, using a home made furnace (temperature reproduced within
2\degr C) in the palaeomagnetic laboratory of the University of
Montpellier. Before the treatment, and after each heating-cooling
cycle, we measured the remanence with a JR5-A spinner magnetometer. To
ensure the reproducibility of the procedure at the same
demagnetization step (pair of heating-cooling cycles and pTRM check),
we always kept the specimens in the same place in the oven.

\subsection{Preliminary selection of palaeointensity data}
Because the interpretation of palaeointensity experiments is
subjective, as many other data interpretations, {\textit{a
    posteriori}} criteria help to ensure the technical quality and
objectivity of the results and their comparison from study to study.
It is noteworthy to point out that except the number of data used for
each individual determination, which has been fixed to four
consecutive points, the other {\textit{a posteriori}} criteria were
not used directly to reject a determination. When only one or two
{\textit{a posteriori}} criteria failed, we kept the determination
choosing the nearest temperature interval for which the corresponding
criteria stay as close as possible from the exclusion bounds fixed.
The data with more than three independent criteria unfulfilled have
been considered as unable to furnish a reliable palaeointensity
estimate and thus have been excluded.\\

\begin{figure}
  \includegraphics{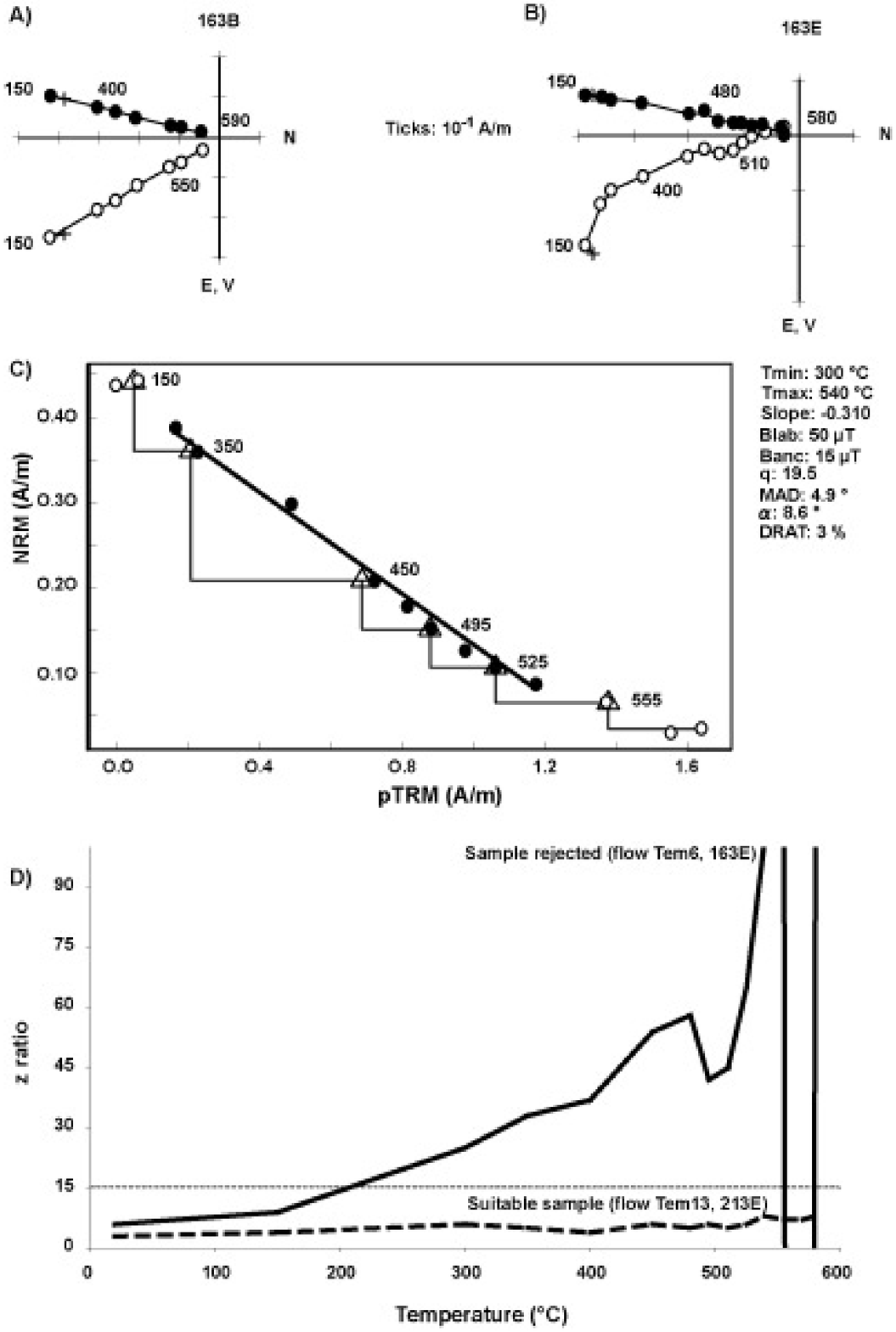}
\caption{Illustration of the a posteriori criteria used to select sample. A) Orthogonal projection of the sister specimen demagnetization, B) Orthogonal projection of the studied
  sample demagnetization, calculated from the palaeointensity
  experiment, C) NRM-TRM diagram of the same specimen, and D)
  Evolution of its z ratio compared to the evolution of the suitable
  sample 213E.}
\label{z}
\end{figure}

\begin{enumerate}
  
\item {$f$ criterion: A classical way of representing palaeointensity
    data is to use the Arai diagram~\cite{Arai63,Nagata63} in which
    the NRM$_T$ remaining at each temperature step T is plotted
    against the pTRM$_T$ acquired in the laboratory field from T to
    room temperature.  The slope of the least squares fit line
    computed from the linear part of the NRM-pTRM diagram gives an
    estimate of the palaeofield strength. Thus criteria have to be
    defined to constrain the determination of this best fit line and
    to quantify its technical quality.\\
    We already fixed the minimum number of successive points used for
    the determinations to four. The other criteria used is the NRM
    fraction $f$ given by the ratio of the NRM lost over the selected temperature
    interval to the total
    NRM~\cite{Coe78}. We consider that a value of 0.3 is the minimum
    NRM fraction for acceptable determination. Note that a quality
    factor $q$~\cite{Coe78,Prevot85} less than 5 can also indicate a
    determination of the palaeofield strength of poor quality.
    However, because $q$ is not independent of $f$, this lead to
    reject the same samples in this study. Thus we did not
    use it as a rejection criterion.\\
    }
\item{Control on vector endpoint diagrams (Zijderveld's plot): Because
    a straight line in the Arai plot can involve more than one
    component, we checked their existence on the Zijderveld's
    projection of the NRM demagnetization computed from the
    palaeointensity experiments. To complete this qualitative
    approach, we quantified the dispersion of the points regarding to
    the best fit line by the maximum angular deviation
    (MAD)~\cite{Kirschvink80} and chose a
    maximum value of 10\degr~for this criterion.\\
    Another quality criterion is given by the angle $\alpha$ between
    the best fit line (anchored to the center of mass) and the vector
    average (anchored to the origin) of the selected data in the
    Zijderveld plot.  If the data correspond to the primary remanence,
    $\alpha$ should be inferior to say 10\degr, in the opposite, they
    are certainly biased by a spurious unknown component.\\
    }
\item {$z$ ratio: The heating remanent magnetizations
    (HRM)~\cite{Calvo02} acquired during the heating procedure under
    the application of a constant weak field in palaeointensity
    experiments, lead to erroneous data. Hopefully, the direction
    along which the laboratory field was applied
    corresponds to the axis of our cylindrical samples (Z axis).
    Therefore, in a Zijderveld plot in sample coordinates, the
    acquisition of HRM appears as a progressive deviation of the
    demagnetization curve in the vertical plane towards the vertical
    axis direction. This is illustrated in Fig.~\ref{z} in which we
    compare the Zijderveld's plot inferred from the
    palaeointensity experiment for a rejected sample with the
    demagnetization curve of its sister specimen obtained during
    the palaeodirection analysis.\\
    An "HRM check" is allowed using the ratio~\cite{Avto99a}:
    \begin{equation}
       \mathrm{z= \frac{HRM_{T}}{NRM_{T}}\times100\%}
    \end{equation}
    where $\mathrm{HRM_{T}}$ and $\mathrm{NRM_{T}}$ are the HRM
    created in the sample and the NRM left in the ChRM direction, at a
    given temperature T, respectively. Similar ratios exist (e.g. R or
    R'~\cite{Coe84}) but the $z$ ratio allows to monitor the evolution
    of alterations during treatment. Calculating this ratio supposes
    that one knows the ChRM direction, which requires
    demagnetization of a sister specimen before the palaeointensity experiment.\\
    To help the interpretation, we defined the upper limit of the
    accepted temperature interval as the one preceding a $z$
    \textgreater 20\%.  However, we extended the interval over this
    limit when the ratio NRM left/TRM gained did not change in the new
    interval. In this case some magneto-chemical transformations are
    present but do not change the estimate. A limitation of the $z$
    ratio is that it depends on the angle between the ChRM and the
    applied field direction (direction of the HRM). The more this
    angle tends to 90\degr, the bigger is the deviation.
    Unfortunately, the tray we used to place the specimens in the oven
    did not permit us to orient their ChRM at 90\degr ~to the applied
    field.  Hence, the $z$ we calculated is a minimum value, thus
    these steps had to be rejected. Another limitation of this
    criterion is that some unexpected fluctuations may appear at high
    temperature when the NRM left is too small. For this reason, we
    completed this approach by observing the evolution of the
    demagnetization on an equal area projection. In Fig.~\ref{z}, the
    $z$-T curve of a rejected sample 163E (Tem06) is compared to the
    one obtained from an accepted sample (213E, flow Tem13).}
  
\item {Difference ratio (DRAT): We performed sliding pTRM checks in
    order to estimate the temperature at which alteration of the
    magnetic minerals begins. Commonly, many authors consider that
    repeated pTRM acquisition at the same temperature steps should
    agree to within 15\%~\cite{Avto99b}.  However, pTRMs acquired
    within low-temperature interval, are rather small and thus a
    reproducibility within a given percentage is difficult to obtain.
    Therefore, we used the difference ratio (DRAT)~\cite{Selkin00},
    which normalizes the maximum difference between repeated pTRM
    steps performed on the temperature interval selected for the
    determination by the length of the corresponding NRM-pTRM segment.
    We fixed a maximum acceptable value of 10\% for this criterion.
    However, this ratio reveals possible alterations on the estimate
    interval only, whereas physical and chemical alterations of the
    heated samples can appear as soon as the first demagnetization
    steps~\cite{Kosterov98b}.  Therefore, we also monitored the lower
    temperature pTRM checks by determining the DRAT from ambient to
    maximum temperature of the interval used to estimate the
    palaeointensity. It is noteworthy that the sample 163E
    (Fig.~\ref{z}), rejected because of its $z$ ratio, has a DRAT of
    only 3\%. Also five samples which give an estimate in
    Table~\ref{tabpi}
    possess a DRAT \textgreater 10\%, thus these two alteration criteria are complementary.\\
    }
  
\item{High temperature pTRM-tail test: Unfortunately, 56\% of the
    samples present a more or less pronounced curvature in their
    NRM-TRM diagrams and some specimens provide two acceptable
    estimates regarding the selection criteria. Because independence
    and additivity laws of pTRM~\cite{Thellier38} are violated for MD
    but maybe also for PSD grains, the determination of the domain
    structure could be a relevant tool to
    choose between the two acceptable palaeointensity estimates.\\
    In case of thermally stable samples, we used the thermomagnetic
    criterion that was first introduced by Bol'shakov \& Shcherbakova
    ~\shortcite{Bolshakov79} to get insight on the domain structure of
    ferrimagnetics. We calculated the parameter $\mathrm{A_{HT}}$
    which is illustrated in Fig.~\ref{tail_hl} and defined by:
    \begin{equation}
      \mathrm{A_{HT}(T_{1},T_{2})=\frac{tail_{HT}[pTRM(T_{1},T_{2})]}{pTRM(T_{1},T_{2})}\times100 \%}
    \end{equation}
    where $\mathrm{pTRM(T_{1},T_{2})}$ is the pTRM measured at room
    temperature gained between T$_1$ and T$_2$ ($\mathrm{T_1 < T_2}$)
    during cooling from the Curie temperature (T$_c$) in a 100$\mu$T
    field, and the $\mathrm{tail_{HT}[pTRM(T_{1},T_{2})]}$ is the part
    of this pTRM not demagnetized when the sample is subsequently
    heated again to T$_1$ and cooled down is zero field.  According to
    the criteria defined by Shcherbakova et
    al.~\shortcite{Shcherbakova00} for $\mathrm{A_{HT}(T_{1},T_{2})}$
    \textless 4\%, the remanent carrier are predominantly SD grains,
    for 4\% \textless $\mathrm{A_{HT}(T_{1},T_{2})}$ \textless
    (15-20)\%, they behave as pseudo-single domain (PSD) grains, and
    for
    $\mathrm{A_{HT}(T_{1},T_{2})}$ \textgreater 20\% they are predominantly MD grains.\\
    We measured the coefficients $\mathrm{A_{HT}(T_{1},T_{2})}$ at
    increasing temperature intervals for one thermally stable sample
    per flow in order to determine the best temperature interval to
    estimate the paleointensity.  Because the heatings were performed
    in air, we repeated the determination of the coefficient
    $\mathrm{A_{HT}(300,T_{room})}$ two times, at the beginning and at
    the end of the treatment, in order to monitor alterations
    appearing during the successive heatings.  Unfortunately, the
    vibrating thermal magnetometer (VTM) we used broke before we could
    measure all the representative samples of each selected flow. The
    results available are reported in Table~\ref{vtm}.  Even though
    all the flows have not been studied, the global trend observed
    confirms that with increasing temperature, the magnetic carrier
    behaviour is more SD-like~\cite{Carvallo03,Shcherbakova00}.  Thus,
    when two coexisting palaeointensity estimates remain possible
    (5 samples), the high temperature pTRM-tail test favors
    acceptance of the higher temperature interval determination,
    as for sample 187E shown in Fig.~\ref{figvtm}.\\
    }
  
\item{Low temperature pTRM-tail test: According to
    Fabian~\shortcite{Fabian01}, the pTRM tail is not a direct measure
    for a sample's tendency to yield a curved Arai plot. The common
    concave-up shape of MD Arai diagrams is rather well accounted for
    by the low temperature tail (Low-T tail) of each pTRM segment
    acquired during cooling from Tc
    ($\mathrm{pTRM_{LT}(T_{1},T_{2})}$).  This Low-T tail is due to
    the grains with unblocking temperature less than blocking
    temperature.  In order to compare the results obtained from the
    two tail tests and validate the choice of the temperature
    intervals for the interpretation, we decided to evaluate the Low-T
    tail of some characteristic samples used during the
    palaeointensity experiment.  We define the coefficient:
    \begin{equation}
    \mathrm{A_{LT}(T_{1},T_{2})=\frac{tail_{LT}[pTRM(T_{1},T_{2})]}{pTRM(T_{1},T_{2})}\times100\%}
    \end{equation}
    where $\mathrm{pTRM(T_{1},T_{2})}$ is the pTRM measured at room
    temperature gained between T$_1$ and T$_2$ ($\mathrm{T_1 < T_2}$)
    in a 50$\mu$T field during cooling from the Curie temperature and
    $\mathrm{tail_{LT}[pTRM(T_{1},T_{2})]}$ is the part of this pTRM
    removed when the sample is subsequently heated to T$_2$ and cooled
    down in zero field (Fig.~\ref{tail_hl}).  This coefficient is
    analogous to the $\mathrm{A_{HT}}$ coefficient defined by
    Shcherbakov et al.~\shortcite{Shcherbakov01b} but the relation
    between the magnitude of the two tails is not clear. We can expect
    from Dunlop \& \"{O}zdemir~\shortcite{Dunlop01} that
    $\mathrm{A_{LT}}$ is equal to $\mathrm{A_{HT}}$ because the
    unblocking temperature distributions are almost symmetric for a
    given blocking temperature, but this may be quite different for
    hundred-degree temperature intervals. In the absence of a
    quantitative study, only the evolution of the
    $\mathrm{A_{LT}}$ ratio for increasing temperature intervals will be considered.\\
    The $\mathrm{A_{LT}}$ coefficients for the (450,350) and (550,450)
    temperature intervals are reported in Table~\ref{l3t}. The main
    result is that $\mathrm{A_{LT}(550,450)}$ is always smaller than
    $\mathrm{A_{LT}(450,350)}$; thus the Low-T tail test again favors
    the higher temperature interval in case of two acceptable
    determinations. In order to estimate the alteration which may
    occur during the required preliminary heating up to T$_c$, the
    untreated samples 187D and 201C have been incorporated in the
    experiment. Comparing their results with the data from the
    corresponding sister samples 187E and 201E, we can observe that,
    even though these samples appear thermally stable (reversible k-T
    curves), $\mathrm{A_{LT}(450,350)}$ is greater for the sample used
    in the palaeointensity estimate and on the contrary
    $\mathrm{A_{LT}(550,450)}$ is greater for the "virgin" sample.
    Nonetheless, for the five samples that allowed two acceptable
    interpretations, we choose the higher temperature interval, which
    yields determination more coherent with the other samples from the
    same flow.  }
\end{enumerate}

\subsection{Palaeointensity results}
\begin{figure}
  \includegraphics{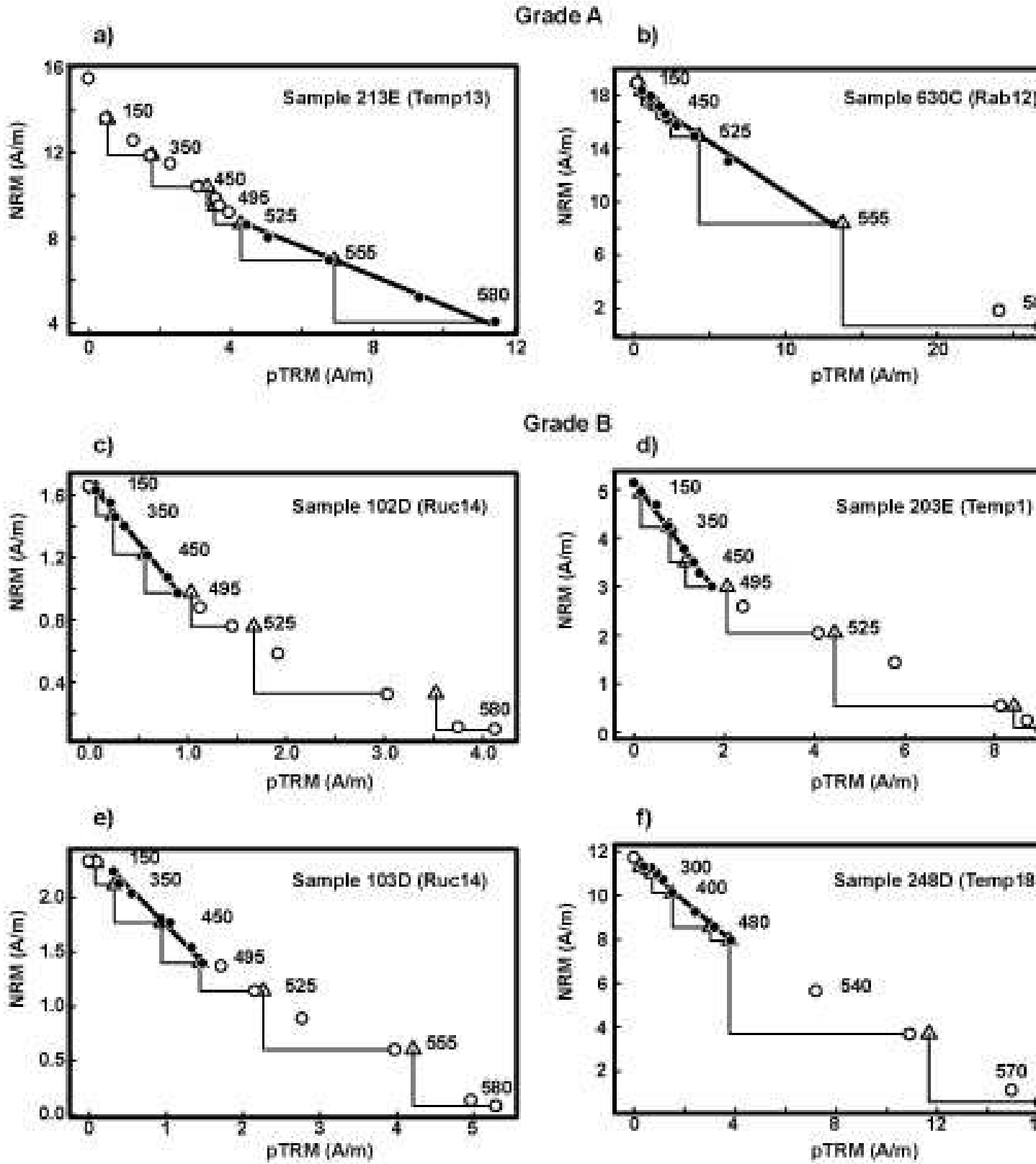}
\caption{Characteristic examples of palaeointensity estimates from this study. a) and b) correspond to grade A
  estimates, c), d), e) and f) to grade B estimates. Black (White)
  circles are the step used (not used) for the palaeointensity
  estimate, the triangles represent the pTRM checks.} \label{figpi}
\end{figure}

Out of the 402 samples from the Mont des Ruches, Mont des
Temp\^{e}tes, and Mont Rabouill\`{e}re sections, 102 have been chosen
using {\textit{a priori}} criteria. After selection following
{\textit{a posteriori}} criteria, 49 samples
furnish a technically acceptable determination (Table~\ref{tabpi}).\\
The quality factor q varies from 1.6 to 54.3 with a mean value of 12.1
for estimates made on a mean of 7 points.  However, some $\alpha$
angles are relatively large, which could be due to an overprint of
viscous origin, even though the samples are not strongly viscous, or
to a secondary component acquired during the treatment, as shown by
the angle $\beta$ between the ChRMs of the sample and its sister
specimen. Although secondary HRM exists, we consider its effect
negligible on the determination, considering the $z$ criterion, the
linearity of the Arai diagram and the evolution of the demagnetization
in an equal area projection. Therefore, as evidenced by the MAD, which
is the most widely satisfied {\textit{a posteriori}} criterion, we are
confident about these interpretations. Moreover, as is commonly done
in palaeointensity studies, we distinguished two types of
determination: class A for the samples which passed successfully all
the criteria and class B for those in which one to three criteria were
unfulfilled. However, as observed with flows Rab12 or Tem6, there is
no single relation between the field estimate and its classification.
Thus, the interpretations made with a few unsatisfied {\textit{a
    posteriori}} criteria can be also considered as reliable.
Fig.~\ref{figpi} shows some characteristic
examples of palaeointensity determinations.\\
We present, in Table~\ref{tabpi}, the mean palaeofield strength with
its associated standard deviation for each individual flow.  Three
consecutive Mont des Ruches flows (Ruc14, 15, and 16) yield a well
defined palaeointensity in the range 35-40 $\mu$T. This value is
comparable with the only reliable flow from the Mont Rabouill\`ere
section (Rab12) and with the flows Tem13 and Tem15 from the Mont des
Temp\^{e}tes. For this last section we also observe three
significantly higher values around 50-60 $\mu$T (Tem1,17 and 18) and
two well defined lower field strengths of only 20.1 and 25.0$\pm$1.7
$\mu$T (Tem6 and 10). Because, the within flow standard deviation
represent at the maximum 15\% of the palaeointensity value (flow
Tem13),
we are quite confident about the reliability of these estimations.\\
We performed in a previous study~\cite{Plenier02}, a quantitative
bootstrap test for a common mean on successive magnetic directions in
order to identify non-independent records. We concluded in particular
that the Ruc14 and Ruc15 flows may represent two contemporaneous
records of the palaeofield (Table~\ref{dir}).  The palaeointensity
estimates obtained for these two successive flows overlap within their
uncertainties and thus strengthen our former interpretation.  However,
because the palaeomagnetic field may remain constant for a relatively
long time~\cite{Love00} we will nevertheless consider in the present
study each of the
12 reliable palaeointensity estimates as distinct records.\\
Concerning the samples 101F, 111F, 201E, 160E, 161F and 229C, for
which pTRM-tail tests were performed, we decided to keep the lower
rather than the higher temperature interval estimates because the
latter was sometimes impossible to define owing to alteration of the
pTRM blocking temperature spectra (101F, 160E, 201E and 229C) or
meaningless compared to the other determinations from the same flow
(111F, 161F).

\section{Discussion}
\subsection{Toward an improvement of palaeointensity determination}
Because of the strict conditions imposed by the commonly used
palaeointensity determination method of Thellier \&
Thellier~\shortcite{Thellier59}, numerous checks are needed to ensure
reliable estimates. For this reason, palaeointensity experiments are
time consuming and at the end many samples do not furnish reliable
determinations. Consequently, the number of reliable estimates is very
low and studies dealing with palaeointensity determinations are
relatively rare.  Then, effective selection followed by careful
palaeointensity determination method are desirable to increase the
productivity of the experiments and largely implement the number of 
palaeointensity estimates in the databases.\\
However, no ideal selection criteria exist to select suitable
specimens and flows for palaeointensity experiments. For example,
Sherbakov et al.~\shortcite{Shcherbakov01b} proposed to use a
pTRM-tail check as a systematic procedure before the experiments to
discriminate the samples with MD grains. The few determinations of
the low and high temperature pTRM-tail performed in the present study
are not convincing since many samples yielding a paleointensity
estimate with good technical quality (Table~\ref{tabpi}) will not have
been kept if we followed Shcherbakov et al.'s
\shortcite{Shcherbakov01b} recommendation. Thus, we do not recommend
to use the thermomagnetic criterion for selection but rather in a case
to case basis to define the best temperature interval for thermally
stable specimens exhibiting two slopes in the NRM/TRM diagram. For the
selection based on k-T curve shapes, the temperature intervals used
for the estimates (c.f. Table~\ref{tabpi}) can change at the flow
scale (flow Tem13 or Tem15) without important deviation in the
palaeofield strength. Therefore, the thermal behaviour of the samples
varies within a flow and this criterion applied to a
supposedly representative sample is not totally effective.\\
A study of each successful sample, at the end of the palaeofield
strength determination procedure, up to the maximum temperature
reached for the palaeointensity estimate would be more interesting.
Whatever the numerous {\textit{a priori}} precautions we took in this
study, only 48\% of the selected samples gave a reliable estimate.
Thus, in the absence of relevant {\textit{a
    priori}} criteria, it is preferable to reduce the preliminary
experiments to the determination of the viscosity index, which is an
already rapid and reliable way to discriminate the specimens disturbed
by a viscous secondary component, followed by a directional analysis,
needed for careful palaeointensity estimates and giving enough
information to discard possibly problematic flows because of large NRM
scatter, poor technical quality of demagnetization curves and so on.\\
As regard to the determination procedure itself, some modifications of
the Thellier's method have been proposed recently
~\cite{Calvo02,Riisager01}.  They include an additional heating at
each demagnetization step in order to control the HRM creation and/or
a ``pTRM tail check''. However, the lack of suitable selection
criteria lead to more failure of the palaeointensity experiment and
these more time consuming procedures are consequently not helpful for
systematic studies. Orienting the samples so that their ChRM is
perpendicular to the direction of the applied field is less time
consuming and sufficient to check for possible HRM creation. Likewise,
at the end of the treatment, complementary experiments like k-T curve
shape determination, extended Thellier's method~\cite{Fabian01} for
thermally stable samples or "pTRM tail checks" can be performed to
ensure the quality and sharpen the reliable estimates, but such
supplementary processes must concern the acceptable samples only.
Thus, good determinations do not necessarily need supplementary
heatings nor burdensome procedures to be performed; this should be
limited only to the problematic samples. Reducing the selection and
placing the control procedures at the end of the treatment is thus a
way to perform reliable palaeointensity estimates more quickly,
without lowering the quality of the determinations.

\subsection{Comparison with previous palaeointensity results}
\begin{figure}
  \includegraphics{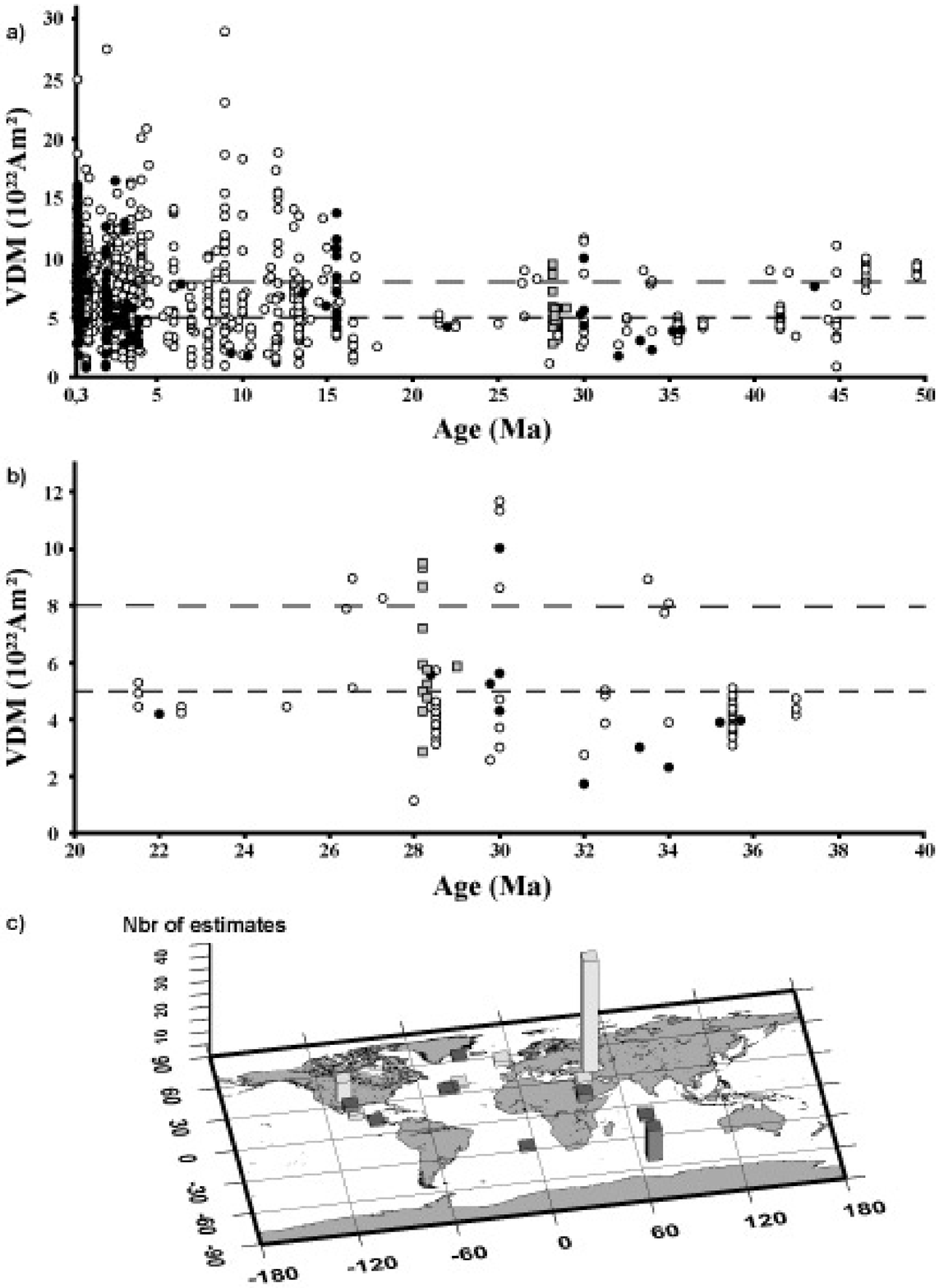}
\caption{a) black(open) circles are selected(unselected) VDMs from the IAGA 2002 updated 
data-set between 0.3 and 50 Ma. Gray
  squares are the data from this study which pass the selection
  criteria. Long (short) dashed line indicates the 0-0.3 Ma (0.3-300
  Ma) mean VDM value. b) same as a) but for the extended Oligocene
  (20-40 Ma) time window. c) Location of the palaeointensity records
  between 20 and 40 Ma. Dark(light) gray column are selected(unselected)
  estimates from the IAGA 2002 updated data-set for the time window
  considered.} \label{figvdm}
\end{figure}

Five Oligocene flows from the Ile Haute section (49.4\degr S,
69.9\degr E) have been processed for palaeointensity study
~\cite{Derder90}. Of the 24 samples analyzed, 11 yielded a
determination, but only one flow provided more than three estimates of
the palaeofield strength.  Moreover, the uncertainty of the mean
palaeointensity is equal to 36\% and consequently, these data cannot
be considered as reliable. The 12 distinct mean flow estimates
presented in this paper are thus the only reliable determinations available for the Kerguelen Archipelago.\\
However, data well distributed on the Earth's surface are needed to
provide a correct idea of the palaeomagnetic field behaviour. It is
then important to pool the Oligocene Kerguelen results with other
reliable determinations already
achieved. For this aim, we used the updated IAGA 2002 palaeointensity database available at the following address:\\
\centerline{ftp://ftp.dstu.univ-montp2.fr/pub/paleointdb/}\\
This database presents estimates obtained with different quality
determinations. However, inclusion of lower quality data leads to
higher average values of the geomagnetic field
\cite{Juarez00,Avto99b}, thus a selection using identical criteria to
those used in this study is needed before we can combine them
meaningfully with the Kerguelen results. For this reason we discarded
estimates based on intermediate polarity flows, obtained with other
methods than the Thellier \& Thellier's ~\shortcite{Thellier59}
associated with pTRM checks, presenting less than 3 individual samples
from the same unit (except for basaltic submarine glasses) and a
standard deviation of the mean $\ge$20\%. Application of these drastic
selection criteria led to the rejection of 660 determinations out of
the 910 estimates contained in the database between 0.3 and 50 Ma
(Fig.~\ref{figvdm}a).  Then we sharpened the selection to a time
window between 20 and 40 Ma, which resulted in 11 suitable estimates
among the 74 initial extended Oligocene records of the data base
(Fig.~\ref{figvdm}a).\\
In order to compare the Kerguelen determinations with the selected
estimates issued from various locations and recording different time,
we calculated the virtual dipole moment (VDM) given by:
\begin{equation}
\mathrm{VDM=\frac{4 \pi R^{3}}{\mu_0} F_{ancient} (1 + 3 \cos^2 \theta)^{-1/2}}
\end{equation}
which corresponds to the moment of a
dipole field producing the estimated palaeointensity $\mathrm{F_{ancient}}$ at the magnetic colatitude
$\theta$. R is the Earth radius. 
Results of the calculations are reported in
Table~\ref{tabpi} and the 12 suitable determinations are plotted in
Fig.~\ref{figvdm}.\\
Even though we performed careful estimates in order to avoid any
possible MD effects in the interpretations, which could lead to
systematic higher determinations, the arithmetic mean obtained with
the Kerguelen data, 6.15$\pm$2.1 10$^{22}Am^2$, is definitely higher
than the other Oligocene estimates already achieved
~\cite{Avto01,Riisager99,Juarez98}. Including Kerguelen determinations
leads to an increase of the Oligocene mean VDM from 4.1$\pm$0.5 to
5.4$\pm$2.3 10$^{22}Am^2$. The new Oligocene mean VDM is closer to the
mean VDM estimated for the 0.3 and 5 Ma interval~\cite{Juarez00} and
higher than the mean VDM defined between 0.3 and 300 Ma
(Fig.~\ref{figvdm}).  Thus these selected Oligocene estimates favor a
relatively stable field between 0.3 and 40 Ma and reinforce the idea
of an exceptionally high recent geomagnetic field strength. However,
the lack of good palaeointensity data didn't allow us to go further in
our interpretation. As illustrated in Fig.~\ref{figvdm}c, the
Oligocene data do not uniformly cover the Earth's surface even though
the Kerguelen results equalize the number of estimates available for
each hemisphere.  More reliable determinations are needed to better
constrain the evolution of the palaeofield with geological time. It is
important to note that the Kerguelen results represent half of the
reliable palaeointensity estimates available for the Oligocene, which
could lead to a bias towards higher values. Moreover, only 3 extended
Oligocene determinations are from normal polarity flows. Thus even if
the 12 new estimates are supposed to be independent, the Oligocene
palaeomagnetic field may not be sufficiently time-averaged.

\section{Conclusion}
We carried out a palaeointensity study on
three Oligocene (28-30 Ma) volcanic sections from the Kerguelen
Archipelago (southern Indian Ocean). It complements the palaeosecular
variation directional study already achieved on the same sections: Mont des Ruches,
Mont des Temp\^{e}tes, and Mont Rabouill\`{e}re
~\cite{Plenier02}. Out of the 57 studied units, we considered
32 suitable for palaeofield strength determination experiments when at
least three of their samples pass the following selection criteria:
angle between NRM and ChRM $\le15\degr$ (or secondary component
quickly demagnetized), NRM dispersion not too large, viscosity index
\textless 5 \%, stable
susceptibility at room temperature after each demagnetization step of
a sister sample and reversible susceptibility in temperature curves.
After preliminary experiments, we studied all samples from 12
favorable flows. We considered only the determinations made with at
least 4 successive steps, a $z$ ratio~\cite{Avto99a} $\le20$\%, a
DRAT~\cite{Selkin00} $\le10$\%, f factors~\cite{Coe78} $\ge0.3$, a MAD
$\le10\degr$~\cite{Kirschvink80} and an angle $\alpha$ between the
best fit line and the vector average $\le10\degr$ to be of good
technical quality. In cases of two acceptable estimates, pTRM tail
tests were used to choose the temperature interval corresponding to
the more SD like behaviour. This careful interpretation of the data
leads to 49 reliable estimates. The VDMs calculated for the 12 flows
vary from 2.78 to 9.47 with an arithmetic mean value of 6.15$\pm$2.1
10$^{22}Am^2$. This study gives the first reliable palaeointensity
estimates from the Kerguelen Archipelago and significantly increases
the number of Oligocene data of comparable quality. The new Oligocene
mean VDM calculated, 5.4$\pm$2.3 10$^{22}Am^2$, is very close to the
0.3-5 Ma value of Juarez \& Tauxe~\shortcite{Juarez00} (5.5$\pm$2.4
10$^{22}Am^2$), and suggests little evolution of the geomagnetic field
between 0.3 and at least 40 Ma. Thus the present moment of the
field (8 10$^{22}Am^2$) is confirmed to be exceptionally high.
However, the lack of reliable data limits the interpretation and we
discuss practical solutions to speed the selection of suitable samples
for careful determination procedures in order to facilitate systematic
studies and increase existing palaeointensity databases with
high-quality estimates.

\begin{acknowledgments}
  We are grateful to the "Institut Polaire Paul Emile Victor" for
  providing all transport facilities and for the support of this
  project. Special thanks to Alain Lamalle, Roland Pagny and all our
  field friends. We thanks Michel Pr\'evot for scientific discussions,
  Thierry Poidras and Liliane Faynot for technical help during k-T,
  VTM and palaeointensity experiments. This work was partially
  supported by CNRS-INSU programme int\'erieur Terre.
\end{acknowledgments}

\bibliographystyle{gji}
\bibliography{GC319}

\newpage

\begin{figure}
  \includegraphics{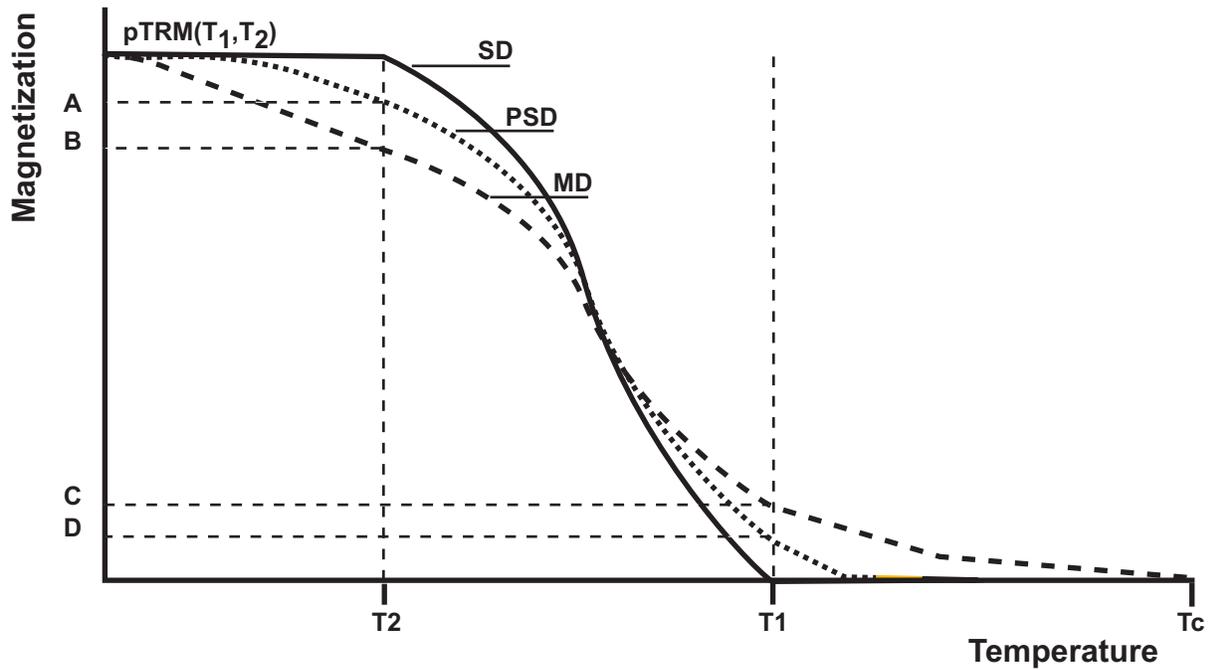}
\caption{Continuous thermal demagnetization of pTRM(T$_1$,T$_2$). The low-temperature pTRM-tail
for PSD (A) or MD (B) grains corresponds to the part of pTRM(T$_1$,T$_2$) removed
at T$_2$, while the high-temperature pTRM-tail corresponds to the part of this pTRM unremoved at T$_1$, 
(C) for MD and (D) for PSD grains. The low and high-temperature pTRM-tails are measured at room 
temperature.}
\label{tail_hl}
\end{figure}

\begin{figure}
  \includegraphics{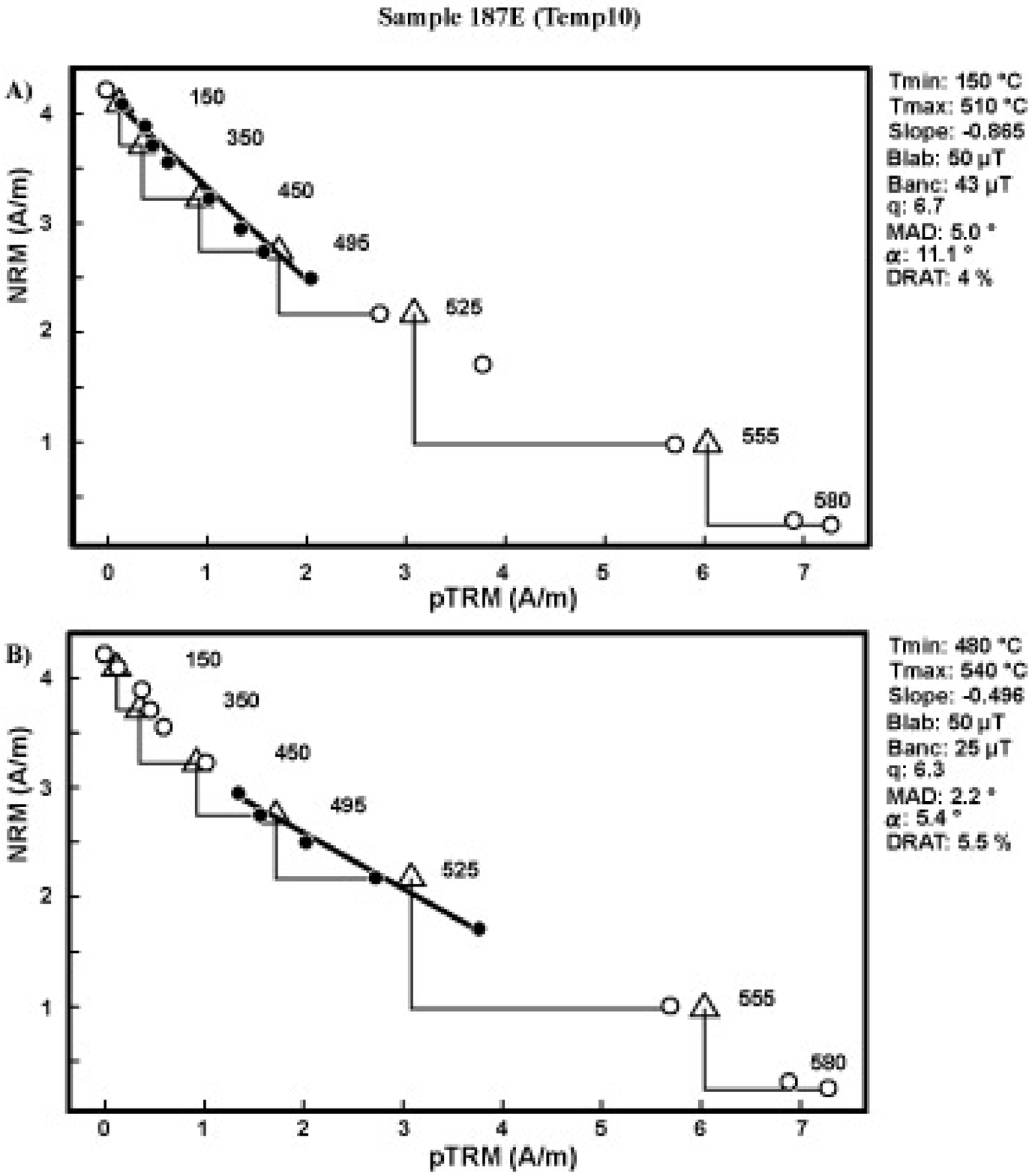}
\caption{Arai diagram for sample 187E (flow Tem10) illustrating a case for which two acceptable 
interpretations are possible. A) The low-temperature interval is rejected regarding the pTRM-tail test
(Table~\ref{vtm}) which favours the high-temperature interval determination shown in B). Black (White) circles are the step used (not used)
  for the palaeointensity estimate, the triangles represent the pTRM checks.} \label{figvtm}
\end{figure}

\newpage

\renewcommand{\tabcolsep}{0.6pc} 
\renewcommand{\arraystretch}{0.6} 

\begin{table*}
 \begin{minipage}{17cm}
 \caption{Cleaned average directions of magnetization of lava flows retained for paleointensity experiments from Plenier et al. (2002).}
 \label{dir}
 \begin{tabular}{lccr@{.}lr@{.}lr@{.}lr@{.}lr@{.}lr@{.}lr@{.}l rc} \\
\hline
 Flow &Chron &n/N &\multicolumn{2}{c}{Inc} &\multicolumn{2}{c}{Dec}
  &\multicolumn{2}{c}{$\alpha _{95}$} &\multicolumn{2}{c}{$\kappa$} &\multicolumn{2}{c}{Plat} &\multicolumn{2}{c}{Plong} &
  \multicolumn{2}{c}{$\nu$\%}&Alt\\
  \hline
  \multicolumn{18}{c}{Mont des Ruches section (48.87$\degr$S, 68.91$\degr$E)}\\
  Ruc16        & C9r&  4/4&   79& 9&  183& 3& 6& 5&  201& 6& -68& 4&  65& 9& 2& 48& 180\\
  Ruc15$^{(1)}$& C9r&  7/7&   74& 5&  159& 2& 3& 9&  244& 8& -73& 1& 105& 3& 4& 60& 160\\
  Ruc14$^{(1)}$& C9r&  7/7&   76& 4&  149& 5& 4& 5&  183& 8& -67& 7& 104& 5& 5& 80& 150\\
  \multicolumn{18}{c}{Mont des Temp\^{e}tes section (48.88$\degr$S, 69.11$\degr$E)}\\
  Tem18        & C9r&  8/8&   60& 7&  177& 0& 3& 5&  248& 7& -82& 5& 231& 6& 2& 73& 172\\
  Tem17        & C9r&  7/7&   62& 0&  186& 0& 4& 0&  224& 8& -83& 0& 287& 7& 3& 36& 160\\
  Tem16        & C9r&  5/7&   64& 2&  178& 1& 4& 2&  328& 6& -86& 8& 224& 6& 5& 57& 152\\
  Tem15        & C9r&  4/4&   64& 5&  188& 1& 2& 8& 1109& 8& -84& 0& 317& 3& 4& 04& 145\\
  Tem13        & C9r&  7/7&   55& 5&  180& 5& 5& 0&  146& 0& -77& 2& 250& 9& 2& 99& 130\\
  Tem10        & C9r&  7/7&   60& 6&  188& 0& 7& 0&   75& 6& -80& 8& 289& 7& 2& 56& 105\\
  Tem6         & C9r&  5/5&   77& 1&  221& 2& 2& 4&  979& 6& -63& 0&  31& 9& 4& 94&  75\\
  Tem1         & C10n.1n& 7/7&  -67& 3&  356& 6& 2& 8&  464& 7&  87& 5& 309& 2& 5& 87&   5\\
  \multicolumn{18}{c}{Mont Rabouill\`{e}re section (49.09$\degr$S, 69.44$\degr$E)}\\
  Rab12        & C10r&    7/7&   68& 2&  140& 2& 4& 2&  207& 0& -64& 8& 138& 7& 1& 80& 150\\
  \hline
  \multicolumn{19}{p{17cm}}{\footnotesize{$^{(\textit{1})}$ indicates flows which have been grouped 
together in Plenier et al.'s (2002) directional analysis. Flows are listed in stratigraphic order
with the youngest on top, oldest on the bottom. Chron correspond to the polarity chrons inferred from
Plenier et al.'s (2002) analysis. n/N is the number of samples analyzed/total number of samples collected.  Inc and
Dec are the mean inclination, positive downward, and the declination east of north, respectively.  $\alpha_{95}$ is the
95\% confidence envelope for the average direction.  $\kappa$ is the precision parameter of Fisher distribution.
Plat/Plong is the latitude/longitude of VGP position, respectively. $\nu$\% is the geometric mean
viscosity index \cite{Thellier44}. Alt is the altitude of the flow in meter.}}\\
\end{tabular}
\end{minipage}
\end{table*}

\newpage

\begin{table*}
 \begin{minipage}{13cm}
 \caption{High-temperature pTRM-tail test}
 \label{vtm}
 \begin{tabular}{lllllll}
\hline
Flow  &Sample &300-T$_{room}$ &400-300\degr C      &500-400\degr C      &550-500\degr C      &300-T$_{room}^{*}$\\
      &       &A$_{HT}$ (B)   &A$_{HT}$ (B) &A$_{HT}$ (B) &A$_{HT}$ (B) &A$_{HT}$\\
\hline
Ruc16 &117B   &17.0           &34.6         &n.d.         &n.d.         &n.d.  \\
Ruc15 &111C   &48.8 (9.7)     &48.1 (11.5)  &30.6 (23.3)  &13.9 (55.5)  &50.7  \\
Ruc14 &101C   &20.8 (11.7)    &29.8 (7.5)   &13.3 (27.9)  &~4.0 (52.9)  &17.3  \\
Ruc10 &076B   &25.2 (40.6)    &33.1 (18.1)  &32.7 (12.9)  &~6.8 (28.4)  &20.3  \\
Tem13 &216C   &14.4 (11.4)    &30.1 (9.9)   &~9.5 (32.4)  &10.3 (46.3)  &21.1  \\
Tem10 &187B   &23.9 (10.4)    &25.5 (10.7)  &11.6 (31.8)  &~5.0 (47.1)  &21.3  \\
      &189B   &29.8 (10.1)    &34.1 (7.3)   &21.7 (19.1)  &~5.3 (63.5)  &26.4  \\
Tem6  &160B   &32.8 (14.9)    &17.4 (16.6)  &28.6 (20.1)  &11.5 (48.4)  &30.7  \\
Tem1  &201B   &12.4 (17.5)    &15.5 (3.0)   &16.6 (21.3)  &~3.7 (58.2)  &~8.8  \\
Rab10 &620D   &15.2 (17.1)    &35.0 (6.7)   &11.4 (34.8)  &~7.6 (41.4)  &16.8  \\
\hline \multicolumn{7}{p{13cm}}
{\footnotesize{
A$_{HT}$ values are the 
relative intensities measured at room temperature of the high temperature pTRM-tail expressed in percent
$\mathrm{A(T_1,T_2) = tail[pTRM(T_1,T_2)]/pTRM(T_1,T_2)}$. 
B values shown in parentheses correspond to the percent of the total pTRM 
(e.g., $\mathrm{\sum_{i}pTRM_i}$) each pTRM(T$_1$,T$_2$) represents;  
nd means not determined.
$^{*}$ is a pTRM-tail test (300-T$_{room}$) repeated at the end of the treatment to control the thermal stability of the sample. 
For
$\mathrm{A_{HT}(T_{1},T_{2})}$ \textless 4\%, the remanent carriers are predominantly SD grains, 
for 4\% \textless $\mathrm{A_{HT}(T_{1},T_{2})}$ \textless (15-20)\%, they present a PSD 
behaviour, and 
for $\mathrm{A_{HT}(T_{1},T_{2})}$ \textgreater 20\% they are predominantly MD grains.}}\\
\end{tabular}
\end{minipage}
\end{table*}

\newpage

\begin{table*}
 \begin{minipage}{8cm}
 \caption{Low-temperature pTRM-tail test}
 \label{l3t}
 \begin{tabular}{llll}
\hline 
Flow &Sample &450-350\degr C  &550-450\degr C\\
     &       &A$_{LT}$ (B) &A$_{LT}$ (B)\\
\hline
Tem18 &251C& 15.7 (9.0)& 2.9 (64.6)\\
Tem16 &237E& ~8.2 (9.3)& 2.0 (74.1)\\
Tem15 &229C& 12.3 (9.4)& 5.1 (59.8)\\
      &230D& ~7.3 (8.2)& 3.1 (43.2)\\
Tem13 &213E& ~6.9 (7.6)& 2.3 (31.8)\\
      &214D& 23.5 (7.7)& 3.9 (18.8)\\
      &215F& ~3.6 (8.1)& 3.4 (22.6)\\
      &217D& ~5.7 (6.5)& 2.9 (43.4)\\
Tem10 &187E& 13.0 (7.0)& 2.3 (65.1)\\
      &187D& ~8.8 (7.2)& 2.4 (54.8)\\
      &189E& 12.9 (5.1)& 2.3 (57.9)\\
Tem6  &161F& ~3.8 (11.7)& 1.5 (42.9)\\
Tem1  &201E& 18.3 (3.3)& 2.2 (72.0)\\
      &201C& 10.1 (6.3)& 2.4 (73.9)\\
\hline \multicolumn{4}{p{8cm}}
{\footnotesize{
$\mathrm{A_{LT}(T_1,T_2)}$ are
the relative intensities measured at room temperature of the part of the pTRM(T$_1$,T$_2$) removed after heating to T$_2$ in zero field.
B values shown in parentheses correspond to the percent of the total TRM (pTRM(580,$T_{room}$)) each pTRM(T$_1$,T$_2$) represents.}}\\
\end{tabular}
\end{minipage}
\end{table*}

\newpage

\renewcommand{\tabcolsep}{0.2pc} 
\renewcommand{\arraystretch}{0.5} 

\begin{center}
\begin{table*}
 \begin{minipage}{20cm}
 \caption{Palaeointensity determinations}
 \label{tabpi}
 \begin{tabular}{llcc@{$\pm$}crccccccccc@{$\pm$}lc}
\hline Flow &Spl &Grade &\multicolumn{2}{c}{Fe$\pm$$\sigma$Fe} &\multicolumn{1}{c}{$\Delta$T} &n &f &g &q &MAD
&$\alpha$ &$\beta$ &DRAT &\multicolumn{2}{c}{\=Fe$\pm$s.d.} &VDM\\
\hline
Ruc16&  115C& B&  34.0&  1.4&  20-400&  5&  0.32&  0.72&  5.5&   6.2&  (11.7)&  7.9&   0.7&    34.8&  0.9&  4.69\\
&       116C& B&  34.2&  2.6& 150-400&  4& (0.22)& 0.58&  1.6&   3.5&    7.9&   5.7&   0.4& \multicolumn{2}{c}{}\\
&       117F& B&  34.8&  2.2& 150-400&  5& (0.23)& 0.62&  2.2&   2.2&    7.5&   3.0&   4.1& \multicolumn{2}{c}{}\\
&       118F& B&  36.3&  1.3&  20-400&  5&  0.32&  0.68&  6.2&   1.7&  (16.6)& 14.8&   3.5& \multicolumn{2}{c}{}\\
Ruc15&  108C& B&  43.0&  2.9&  20-400&  5& (0.25)& 0.73&  2.7&   4.2&    8.8&   7.8&   0.4&    40.0&  4.3&  5.69\\
&       110E& B&  37.8&  1.1& 350-540&  8&  0.37&  0.81& 10.2&   4.4&  (19.8)& 16.7&   4.1& \multicolumn{2}{c}{}\\
&       111F& B&  45.0&  1.1&  20-400&  6& (0.23)& 0.79&  4.9&   4.6&   11.2&  10.8&   1.7& \multicolumn{2}{c}{}\\
&       114D& B&  34.2&  0.9& 150-480&  6&  0.53&  0.79& 15.9&   1.8&  (16.4)& 11.0&   5.2& \multicolumn{2}{c}{}\\
Ruc14&  101F& B&  38.6&  1.1&  20-480&  8&  0.37&  0.81& 10.8&   4.1&  (15.7)& 12.9&   2.7&    37.2&  2.7&  5.18\\
&       102D& B&  39.6&  1.0& 150-495&  7&  0.39&  0.80& 12.6&   2.5&  (12.2)& 10.5&   2.4& \multicolumn{2}{c}{}\\
&       103D& B&  33.5&  1.8& 300-495&  6&  0.33&  0.75&  4.5&   2.2&  (12.1)&  5.9&   8.1& \multicolumn{2}{c}{}\\
\\
Tem18& 245D& A&  65.4&  2.1&  20-510&  9&  0.44&  0.86& 11.8&  4.2&     8.2&   2.6&   5.5&    55.0&  6.9&  9.30\\
&       246C& B&  43.9&  0.8& 300-495&  6& (0.27)& 0.78& 11.5&  2.7&   (18.2)& 10.1&   7.6& \multicolumn{2}{c}{}\\
&       248D& B&  51.6&  1.8& 150-500&  8& (0.29)& 0.82&  7.2&  5.7&   (13.0)& 10.0&   4.2& \multicolumn{2}{c}{}\\
&       249D& A&  53.0&  4.2&  20-400&  5&  0.36&  0.71&  3.2& 2.6&     7.3&   1.9&   0.1& \multicolumn{2}{c}{}\\
&       251C& A&  54.8&  1.5& 150-525&  9&  0.51&  0.85& 15.8&  3.2&     4.9&   3.9&   4.1& \multicolumn{2}{c}{}\\
&       252D& B&  61.2&  1.6&  20-495&  8&  0.52&  0.81& 16.0&  3.0&   (10.1)&  6.4& (13.8)&\multicolumn{2}{c}{}\\
Tem17& 238D& B&  48.4&  2.5& 150-495&  7& (0.28)& 0.82&  4.5&  3.4&   (14.7)& 11.5&   5.7&    51.8&  6.8&  8.62\\
&       239D& B&  58.8&  2.1& 150-480&  6& (0.25)& 0.78&  5.4&  5.1&   (19.4)& 14.0& (10.1)&\multicolumn{2}{c}{}\\
&       240E& B&  39.2&  0.9& 150-500&  8&  0.40&  0.79& 13.1&  5.1&   (11.1)&  5.7&   5.3& \multicolumn{2}{c}{}\\
&       241D& B&  56.6&  1.1& 150-495&  7&  0.36&  0.82& 15.1&  3.3&   (12.5)&  7.2&   6.3& \multicolumn{2}{c}{}\\
&       243E& B&  57.3&  3.5& 150-450&  5& (0.23)& 0.73&  2.7& 4.9&   (26.3)& 27.5&   6.6& \multicolumn{2}{c}{}\\
&       244E& B&  50.3&  0.9& 150-510&  8&  0.35&  0.84& 16.0&  5.0&   (10.1)&  5.2&   2.2& \multicolumn{2}{c}{}\\
Tem16& 232D& A&  35.2&  0.8&  20-540& 11&  0.54&  0.87& 20.8&  2.8&     8.0&   1.0&   4.2&    30.5&  3.8&  4.95\\
&       233C& B&  30.4&  0.4& 350-540&  8&  0.69&  0.84& 40.5&  2.3&     2.5&   1.9& (10.7)&\multicolumn{2}{c}{}\\
&       234E& B&  24.6&  0.4& 300-540&  9&  0.73&  0.84& 35.3&  3.4&     1.3&   3.0& (15.1)&\multicolumn{2}{c}{}\\
&       237E& A&  31.6&  0.6& 350-540&  6&  0.54&  0.66& 18.1&  2.5&     1.5&   4.3&   2.3& \multicolumn{2}{c}{}\\
Tem15& 228D& A&  34.3&  0.8& 480-580&  8&  0.76&  0.76& 25.3&  3.1&     1.8&   2.6&   7.7&    36.6&  1.6&  5.89\\
&       229C& B&  38.2&  1.7& 150-495&  7&  0.30&  0.82&  5.4&  7.2&   (17.5)& 13.4&   2.1& \multicolumn{2}{c}{}\\
&       230D& A&  37.1&  0.4& 480-580&  6&  0.78&  0.71& 54.3&  1.6&     0.3&   5.1&   5.8& \multicolumn{2}{c}{}\\
Tem13& 213E& A&  32.8&  1.0& 525-580&  5&  0.41&  0.72&  9.8&  2.2&     0.9&   2.8&   2.0&    39.7&  5.7&  7.18\\
&       214D& A&  34.7&  1.7&  20-555& 12&  0.48&  0.85&  8.1&  3.5&     5.8&   6.5&   2.5& \multicolumn{2}{c}{}\\
&       215F& A&  36.6&  1.5&  20-555& 12&  0.58&  0.85& 12.4&  5.1&     2.2&   4.6&   2.4& \multicolumn{2}{c}{}\\
&       216F& A&  40.0&  1.9&  20-500&  9&  0.52&  0.85&  9.1&  7.8&     8.5&   4.0&   5.4& \multicolumn{2}{c}{}\\
&       217D& A&  45.9&  3.3&  20-555& 12&  0.74&  0.85&  8.7&  1.9&     3.3&   2.7&   3.2& \multicolumn{2}{c}{}\\
&       218C& A&  48.3&  3.1&  20-555& 12&  0.50&  0.82&  6.3&  3.5&     6.4&   7.3&   5.2& \multicolumn{2}{c}{}\\
Tem10& 187E& A&  24.8&  1.0& 480-540&  5&  0.35&  0.71&  6.3&  2.2&     5.4&   2.5&   5.5&    25.0&  1.7&  4.23\\
&       188E& A&  27.2&  1.4& 510-555&  4&  0.47&  0.53&  4.8&  4.1&     4.7&   3.2&   5.1& \multicolumn{2}{c}{}\\
&       189E& A&  23.0&  0.5& 525-570&  4&  0.66&  0.62& 17.5&  3.4&     1.1&   6.3&   6.1& \multicolumn{2}{c}{}\\
Tem6&  160E& B&  21.4&  0.5&  20-500&  9&  0.57&  0.83& 21.4&   9.4&  (14.9)&  6.7&   6.0&    20.1&  1.7&  2.78\\
&       161F& A&  21.2&  0.9& 150-525&  9&  0.53&  0.85& 10.6&   3.6&    6.0&   2.8&   4.0& \multicolumn{2}{c}{}\\
&       162E& B&  17.8&  0.5&  20-525& 10&  0.52&  0.80& 16.7& (10.6)& (14.0)&  6.1&   5.5& \multicolumn{2}{c}{}\\
Tem1&  201E& B&  55.9&  4.9& 150-480&  7&  0.31&  0.82&  2.9&   6.1&  (16.9)& 13.2&   1.8&    61.0&  3.6&  9.47\\
&       203E& B&  62.9&  2.6&  20-495&  8&  0.42&  0.84&  8.6&   3.8&  (11.0)&  9.8&   7.3& \multicolumn{2}{c}{}\\
&       205E& B&  64.1&  3.2& 150-510&  8&  0.35&  0.82&  5.7&   4.7&    9.9&   6.6& (13.1)&\multicolumn{2}{c}{}\\
\\
Rab12&  628B& A&  35.5&  0.6& 225-500&  7&  0.53&  0.77& 24.4&  3.3&     5.7&   4.3&   5.2&    37.7&  2.4&  5.76\\
&       629B& A&  38.0&  2.0& 150-480&  7&  0.54&  0.75&  7.7&  2.1&     8.1&   5.2&   7.1& \multicolumn{2}{c}{}\\
&       630C& A&  41.4&  1.1& 150-555& 11&  0.57&  0.72&  7.8&  2.2&     5.8&   6.7&   1.5& \multicolumn{2}{c}{}\\
&       631C& B&  35.8&  1.6& 150-450&  5&  0.30&  0.68&  4.5& 5.3&   (16.1)&  11.9&   3.3& \multicolumn{2}{c}{}\\
\hline\\
\multicolumn{17}{p{16cm}}{\footnotesize{\textit{Grade}: classification based on the number of {\textit{a posteriori}}
criteria checked (A all, B except at least one evidenced with parenthesis); \textit{Fe$\pm$$\sigma$Fe}: individual
palaeofield strength determination (in $\mu$T) and standard error associated~\cite{Prevot85}; \textit{$\Delta$T}:
temperature interval of determination; \textit{n}: number of consecutive points used for the determination; \textit{f},
\textit{g} and \textit{q}: fraction of NRM, gap factor and quality factor, respectively~\cite{Coe78,Prevot85};
\textit{MAD}: maximum angular deviation; \textit{$\alpha$}: angle between ChRM and origin; \textit{$\beta$}: angle
between ChRMs of the sample and its sister specimen~\cite{Plenier02}; \textit{DRAT}: difference ratio
~\cite{Selkin00}; \textit{\=Fe$\pm$s.d.}: mean palaeofield strength of the flow and the standard deviation associated;
\textit{VDM}: virtual dipole moment (*$10^{22}Am^{2}$).}}\\
\end{tabular}
\end{minipage}
\end{table*}
\end{center}

\end{document}